\documentclass[a4paper,twocolumn,superscriptaddress]{revtex4-1}

\usepackage{graphicx}
\usepackage{bm}
\begin{document}

\title{Comment on ``Evidence for dark matter in the inner Milky Way''}

\author{Stacy McGaugh}
\affiliation{Department of Astronomy, Case Western Reserve University, 10900 Euclid Avenue, Cleveland, OH 44106, USA}
\author{Federico Lelli}
\affiliation{Department of Astronomy, Case Western Reserve University, 10900 Euclid Avenue, Cleveland, OH 44106, USA}
\author{Marcel Pawlowski}
\affiliation{Department of Astronomy, Case Western Reserve University, 10900 Euclid Avenue, Cleveland, OH 44106, USA}
\author{Garry Angus}
\affiliation{Department of Physics and Astrophysics, Vrije Universiteit Brussel, Pleinlaan 2, Brussels, 1050 Belgium}
\author{Olivier Bienaym\'e}
\affiliation{Observatoire Astronomique de Strasbourg, Universit\'e de Strasbourg, CNRS, UMR 7550, 11 rue de l'Universit\'e, F-67000 Strasbourg, France}
\author{Joss Bland-Hawthorn}
\affiliation{Sydney Institute for Astronomy, School of Physics A28, University of Sydney, NSW 2611, Australia}
\author{Erwin de Blok}
\affiliation{Netherlands Institute for Radio Astronomy (ASTRON), Postbus 2, 7990 AA, Dwingeloo, The Netherlands}
\affiliation{Astrophysics, Cosmology and Gravity Centre, Department of Astronomy, University of Cape Town, Private Bag X3, Rondebosch 7701, South Africa}
\affiliation{Kapteyn Astronomical Institute, PO Box 800, NL-9700 AV Groningen, the Netherlands}
\author{Benoit Famaey}
\affiliation{Observatoire Astronomique de Strasbourg, Universit\'e de Strasbourg, CNRS, UMR 7550, 11 rue de l'Universit\'e, F-67000 Strasbourg, France}
\author{Filippo Fraternali}
\affiliation{Department of Physics and Astronomy, University of Bologna, viale Berti Pichat 6/2, I-40127 Bologna, Italy}
\affiliation{Kapteyn Astronomical Institute, PO Box 800, NL-9700 AV Groningen, the Netherlands}
\author{Ken Freeman}
\affiliation{RSAA Australian National University, Mount Stromlo Observatory, Cotter Road, Weston Creek, Canberra, ACT 72611, Australia}
\author{Gianfranco Gentile}
\affiliation{Sterrenkundig Observatorium, Universiteit Gent, Krijgslaan 281, B-9000 Gent, Belgium}
\affiliation{Department of Physics and Astrophysics, Vrije Universiteit Brussel, Pleinlaan 2, 1050 Brussels, Belgium}
\author{Rodrigo Ibata}
\affiliation{Observatoire Astronomique de Strasbourg, Universit\'e de Strasbourg, CNRS, UMR 7550, 11 rue de l'Universit\'e, F-67000 Strasbourg, France}
\author{Pavel Kroupa}
\affiliation{Helmholtz-Institut f\"ur Strahlen- und Kernphysik, Universit\"at Bonn, Nussallee 14–16, D-53115 Bonn, Germany}
\author{Fabian L\"ughausen}
\affiliation{Helmholtz-Institut f\"ur Strahlen- und Kernphysik, Universit\"at Bonn, Nussallee 14–16, D-53115 Bonn, Germany}
\author{Paul McMillan}
\affiliation{Lund Observatory, Box 43, SE-221 00 Lund, Sweden}
\author{David Merritt}
\affiliation{School of Physics and Astronomy and Center for Computational Relativity and Gravitation, Rochester Institute of Technology, 84 Lomb Memorial Drive, Rochester, NY 14623, USA}
\author{Ivan Minchev}
\affiliation{Leibniz-Institut fŸr Astrophysik Potsdam (AIP) An der Sternwarte 16, 14482 Potsdam, Germany}
\author{Giacomo Monari}
\affiliation{Observatoire Astronomique de Strasbourg, Universit\'e de Strasbourg, CNRS, UMR 7550, 11 rue de l'Universit\'e, F-67000 Strasbourg, France}
\author{Elena D'Onghia}
\affiliation{Department of Astronomy, University of Wisconsin, 2535 Sterling Hall, 475 North Charter Street, Madison, WI 53076, USA}
\author{Alice Quillen}
\affiliation{Department of Physics and Astronomy, University of Rochester, Rochester, NY 14627, USA}
\author{Bob Sanders}
\affiliation{Kapteyn Astronomical Institute, PO Box 800, NL-9700 AV Groningen, the Netherlands}
\author{Jerry Sellwood}
\affiliation{Department of Physics and Astronomy, Rutgers University, 136 Frelinghuysen Road, Piscataway, NJ 08854, USA}
\author{Arnaud Siebert}
\affiliation{Observatoire Astronomique de Strasbourg, Universit\'e de Strasbourg, CNRS, UMR 7550, 11 rue de l'Universit\'e, F-67000 Strasbourg, France}
\author{Hongsheng Zhao}
\affiliation{SUPA, University of St Andrews, St Andrews, KY16 9SS, UK}

\begin{abstract}
This is a brief rebuttal to arXiv:1502.03821, which claims 
to provide the first observational proof of dark matter interior to the solar circle.
We point out that this result is not new, and can be traced back at least a quarter century.
\end{abstract}

\maketitle

\date{\today}

We were surprised to read the recent Letter by Iocco et al.\ \cite{iocco} claiming new evidence for dark matter (DM) in the inner Milky Way.
Bright spirals like the Milky Way can often be described as maximum disks\cite{N3109orig}: 
the observed rotation velocity in the inner regions can be explained by the observed baryons.
Quantitatively, `inner' here means radii within
2.2 disc scale lengths\cite{S97}, about 5 kpc in the Milky Way\cite{bovyrix}.
While there may indeed be some DM at small radii, the need for it only becomes clear farther out.  

The innocent reader of Iocco et al.\ would get the impression that the data require lots of DM in the inner Milky Way.
Indeed, they claim to have `obtained for the first time a direct observational proof of the presence of dark matter 
in the innermost part of the Milky Way'\footnote{http://perma.cc/JJ4S-G42D}.  
This assertion is not warranted by their analysis.

The innermost part of the Galaxy (radii $< 2.5$ kpc) is not probed at all.  
This region is characterized by non-circular motions due to the stellar bar\cite{Zhao96}.
Bars are the natural consequence of disk self-gravity\cite{OP73}, and are suppressed if there is too much DM.

Figure~\ref{MWfig} reproduces Figure~2 of Iocco et al.\ on a linear scale.
We show the most precise rotation curve data\cite{Clemens,MGD} and their range of baryonic mass models. 
These accommodate both sub-maximal and fully maximal disk configurations. 
The necessity of DM only becomes clear beyond 6 kpc, where the rotation velocity exceeds the upper
bracket of the baryonic contribution.  This result is not new, and follows for any plausible value of the
circular speed and Galactocentric distance of the Sun\cite{BT87}.

\begin{figure*}[t]
\includegraphics[width=5in]{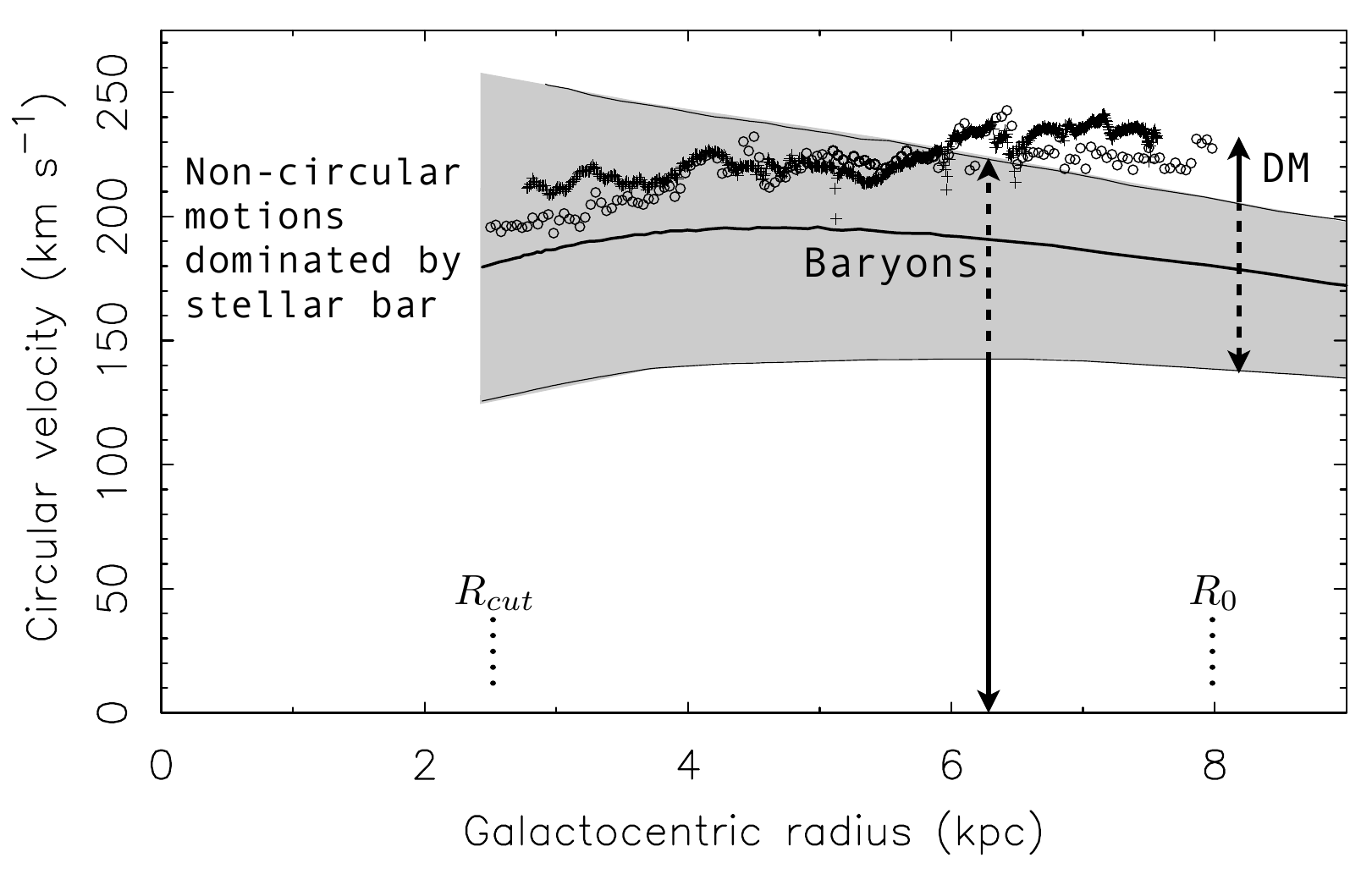}
\caption{The Milky Way rotation curve from the receding\cite{Clemens} (circles) and approaching\cite{MGD} (crosses) sides
of the Galaxy.  Measurement errors are comparable to the symbol size.  
Differences between the two sides may stem from asymmetry in the mass distribution (e.g., spiral arms).
The bracketing baryonic mass model of Iocco et al.\ is shown as the grey band. 
Dark matter is inferred where the observed velocity exceeds this band.
\label{MWfig}}
\end{figure*}

Iocco et al.\ have combined data from many different sources, as have 
others\cite{S97,bovyrix,Zhao96,Clemens}. Great care must be taken in doing this, as
the differences between astronomical datasets are often systematic.  
One cannot simply combine them to obtain a statistically meaningful estimate of the uncertainty. 

Considerable confusion may stem from the use of the term `inner'.
The Sun's orbit encompasses roughly 90\% of the stellar mass\cite{bovyrix}.  
By this standard, we live in the outskirts of the Galaxy. 
That some DM is needed interior to the Solar circle is neither surprising nor new\cite{BT87}. 

\noindent \textbf{Note Added:} Similar concerns to ours have also been expressed by Durazo et al.\ \cite{DHM_hellno}.
Iocco et al.\ have replied to our points above in \cite{wrongagain}.  They make three basic points.

First, they address the confusion over the meaning of the word `inner.' 
Apparently, to them, the Milky Way is not the luminous band that crosses the sky, 
but rather the putative dark matter halo in which it resides.  
Certainly we agree that these two distinct entities should not be confused.  
They do use the term `innermost', which appears in the third sentence of their abstract \cite{iocco}.  
Innermost, by definition, encompasses the region $R \rightarrow 0$, not merely $R > 2.5$ kpc
or the ring $6 < R < 8$ kpc immediately interior to the solar circle where their analysis requires DM.
For an assessment of the possible DM content of the innermost Milky Way ($R < 2.2$ kpc), see \cite{PWGMV15}.

Second, they claim to have made a thorough discussion of the uncertainties in the supplementary discussion.
This is true.  The real issue, however, is that systematic uncertainties dominate over random errors.
The latter may be well quantified, but the former are not.  This is a common circumstance in astronomy, as we cannot
control the conditions of the laboratory.  It is simply not possible to obtain a statistically meaningful error distribution 
in the fashion presented.

Third, they mistakenly presume that we confuse DM at the solar circle with that interior to it.
It is they who are missing a simple point.  Rotation curves are approximately flat.  The portion of the rotation curve
that can be attributed to stars declines at radii beyond a peak at 2.2 scale lengths of the stellar disk.
This divergence requires DM at all radii beyond the peak.  Mass models \cite{M08} that are consistent with the observed
distribution of stars inevitably have this peak interior to the solar radius.  The need for DM interior to the
sun follows directly.

Iocco et al.\ \cite{wrongagain} persist in asserting that their work \cite{iocco} is novel, stating that it provides
`the first direct observational proof of dark matter inside the solar circle.'
Yet the need for DM interior to the solar circle was observationally demonstrated long ago.
Fig.\ 2 reproduces Fig.\ 2 from a 1988 paper \cite{SellwoodSanders} on precisely this topic.
This early work shows that, even for a maximal disk, the need for DM sets in around 6 kpc.  
This result is, in essence, identical to that of Iocco et al.

\begin{figure*}[t]
\includegraphics[width=5.5in]{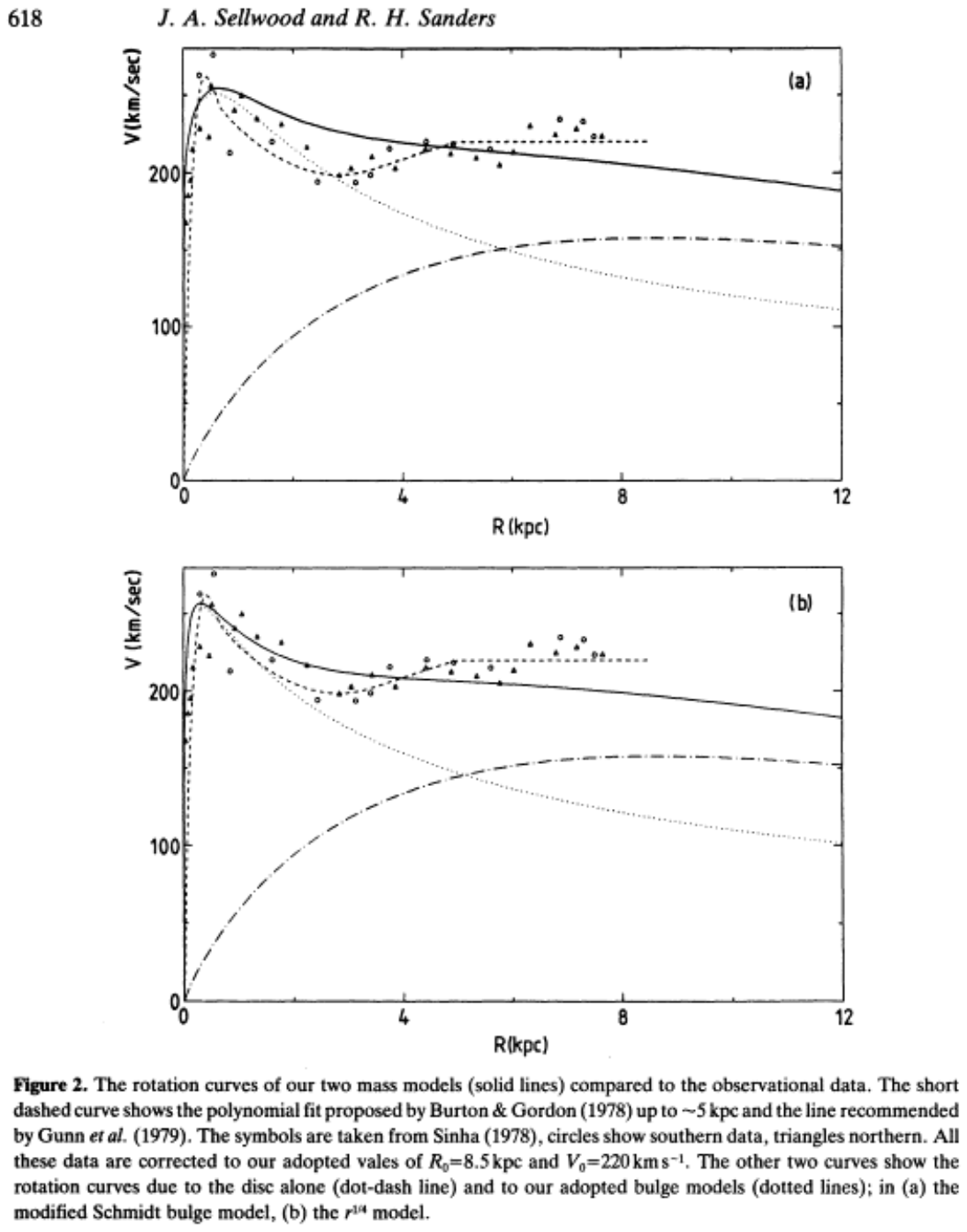}
\caption{Fig.~2 from the 1988 paper `A maximum disc model for the Galaxy' \cite{SellwoodSanders} showing the
need for DM interior to the solar circle.
\label{MWfig}}
\end{figure*}

Iocco et al.\ \cite{wrongagain} imply that their `bracket' of all possible Milky Way morphologies is the first `data-driven'
assessment of the baryonic component.  But this is exactly what maximum disk is: a data-driven upper limit on the baryonic component.
This data-driven assessment appeared in the literature over a quarter century ago, so
the claim of Iocco et al.\ to be the first to demonstrate the need for DM inside the solar circle is factually incorrect.

\bibliography{NatPhysComm_abstract}

\end{document}